\documentclass[prl,twocolumn,superscriptaddress,nofootinbib,
aps,floatfix]{revtex4}

\usepackage{amsthm}
\usepackage{amsmath}   
\usepackage{amssymb}  
\usepackage{amsfonts}   
\usepackage{graphicx}

\newtheorem*{remark}{Note}
\newtheorem{result}{Result}

\let\oldenumerate\enumerate
\renewcommand{\enumerate}{
  \oldenumerate
  \setlength{\itemsep}{1pt}
  \setlength{\parskip}{0pt}
  \setlength{\parsep}{0pt}
}

\let\olditemize\itemize
\renewcommand{\itemize}{
  \olditemize
  \setlength{\itemsep}{1pt}
  \setlength{\parskip}{0pt}
  \setlength{\parsep}{0pt}
}

\allowdisplaybreaks[2]  %allow page breaks in multi-line eqns, [1]=as little as poss; ... [4]=most permissive
\usepackage{pifont}% http://ctan.org/pkg/pifont
\usepackage{bm} %boldmath
\usepackage{mathrsfs}   % Weinberg-esque letters 
\usepackage{dsfont}  %doublestruck math = \mathds
\usepackage{ytableau}   % Young diagrams
       
\usepackage{float}
\usepackage{hhline}
\usepackage{mathtools}

\graphicspath{{figures/}}	 % set directory for figures

\usepackage{array} % ??
\usepackage{multirow} % multiple row elements in a table
\usepackage{booktabs} % more control over tables
\usepackage{nicefrac} % fractions in tables,
\usepackage{stfloats}  % ??
\usepackage{xcolor}   % \color{...}, colored text
\definecolor{aqua}{rgb}{0.0, 1.0, 1.0}

\usepackage{url} % ??
\usepackage{verbatim} %??

\def\bar{\overline}
\def\til{\widetilde}
\def\hat{\widehat}
\def\del{{\partial}}
\def\delb{{\bar\del}}
 
\def\vev#1{{\langle{#1}\rangle}} 
\newcommand{\beq}{\begin{equation}}
\newcommand{\eeq}{\end{equation}}
\newcommand{\bpmat}{\begin{pmatrix}}
\newcommand{\epmat}{\end{pmatrix}}
\def\bmat{\begin{matrix}}
\def\emat{\end{matrix}}
\newcommand{\bsmat}{\begin{smallmatrix}}
\newcommand{\esmat}{\end{smallmatrix}}

\def\^{\wedge}

\def\U{{\rm\, U}}

\def\Sp{{\rm\, Sp}}

\def\msC{{\mathscr C}}
\def\msD{{\mathscr D}}
\def\msM{{\mathscr M}}
\def\msP{{\mathscr P}}

\def\C{\mathbb{C}} 
\def\tCs{\til{\C^*}}

\def\N{\mathbb{N}} 

\def\Q{\mathbb{Q}} 
\def\R{\mathbb{R}} 
 
\def\Z{\mathbb{Z}}

\def\kb{{\bar k}}
\def\bl{{\boldsymbol\ell}}

\def\bp{{\bf p}}

\def\bq{{\bf q}}

\def\bs{{\bf s}}

\def\bu{{\bf u}}

\def\bv{{\bf v}}

\def\cE{{\mathcal E}}

\def\cN{{\mathcal N}}
\def\cO{{\mathcal O}}

\def\cV{{\mathcal V}}

\def\g{{\gamma}}

\def\d{{\delta}}
\def\D{{\Delta}}

\def\ve{\varepsilon}

\def\l{{\lambda}}

\def\m{{\mu}}

\def\s{{\sigma}}
\def\bs{{\boldsymbol\s}}

\def\S{{\Sigma}}

\def\F{{\Phi}}
\def\vf{{\varphi}}

\def\SpDrZ{{\Sp_D(2r,\Z)}}

\begin{document}
\onecolumngrid

\twocolumngrid
%TITLE PAGE

\title{Scaling dimensions of Coulomb branch operators\\ 
of 4d N=2 superconformal field theories}

\author{Philip C. Argyres}
%\email[]{\texttt{philip.argyres@gmail.com}}
\affiliation{University of Cincinnati,
Physics Department, PO Box 210011, Cincinnati OH 45221}
\affiliation{California Institute of Technology, Walter Burke Institute for Theoretical Physics, Pasadena CA 91125}

\author{Mario Martone}
%\email[]{\texttt{mariomartone@utexas.edu}}
\affiliation{University of Texas, Austin,
Physics Department, Austin TX 78712}

\begin{abstract}
Under reasonable assumptions about the complex structure of the set of singularities on the Coulomb branch of $\cN=2$ superconformal field theories, we present a relatively simple and elementary argument showing that the scaling dimension, $\D$, of a Coulomb branch operator of a rank $r$ theory is allowed to take values in a finite set of rational numbers
\begin{align}
\D \in \left\{ \frac{n}{m} \ \bigg\vert\ n,m\in\N,\ 0<m\le n,\ 
\gcd(n,m)=1,\ \vf(n) \le 2r \right\}
\nonumber
\end{align}
where $\vf(n)$ is the Euler totient function.   The maximal dimension grows superlinearly with rank as $\D_\text{max} \sim r \ln\ln r$.  This agrees with the recent result of Caorsi and Cecotti.
\end{abstract}

\maketitle
%\flushbottom

%\section{Introduction \& Outline}

In the quest for a better analytic understanding of the strong-coupling behavior of quantum field theories (QFTs), conformal field theories stand out both as especially symmetric examples and for their usefulness in organizing the space of QFTs as fixed points of their renormalization group flows.  A fundamental piece of information about a conformal field theory is the spectrum of scaling dimensions of its local operators.  Superconformal field theories (SCFTs) with $\cN=2$ extended supersymmetry in 4 dimensions form an especially interesting class since they show a rich variety of phenomena and also are amenable to an exact analysis of observables using a wide array of techniques.  The technique which will interest us here is one of the oldest, going back to the work of \cite{sw1,sw2}, which uses the existence of a moduli space of vacua whose low energy degrees of freedom include photons --- the Coulomb branch (CB) of the moduli space.  The structure imposed on the geometry of the CB by the discrete group of electric-magnetic duality transformations of the photon fields is sufficiently rigid to permit the exact computation of some observables even at strong coupling.

We will show here how these techniques can be used to compute the exact spectrum of scaling dimensions of certain local operators associated to the CB of any $\cN=2$ SCFT.  The significance of this result is that it is independent of any constructions in terms of RG flows from other known SCFTs or from certain string constructions.  The argument presented here, though somewhat intricate, is based on elementary considerations involving the topology of the locus of metric singularities on the CB, the electric-magnetic (EM) duality monodromies along certain symmetry orbits in the CB, properties of unipotent elements of the EM duality group, and positivity of the metric on the CB.  At its core, this argument is a straightforward generalization to arbitrary rank of the results obtained for rank 2 in \cite{Argyres:2018zay}.  Most of the ideas needed here are explained there in detail, allowing us to keep the presentation short and to stress the main physical results.

One interesting result is the asymptotic behavior of the spectrum of dimensions at large rank.  This is related to the large-rank behavior of the conformal central charges via a sum rule and can be compared to large-N results via the AdS/CFT correspondence.

\begin{remark}
When this paper was completed, \cite{Caorsi:2018zsq} appeared with large overlap with our work. Our results are in perfect agreement with theirs.
\end{remark}

\subsection{Elements of Coulomb branch geometry}

A CB is a moduli space of vacua of a 4d $\cN=2$ supersymmetric QFT which has unbroken $\cN=2$ supersymmetry and an IR $\U(1)^r$ gauge symmetry.  Its regular points are vacua where all the massless degrees of freedom are neutral under $\U(1)^r$.  The vevs of the complex scalars of the $r$ free massless vector multiplets at these vacua are complex coordinates  $\bu\in\C^r$ on the CB.  The leading terms of the vector multiplet effective action and the central charges as functions of $\bu$ describe a special K\"ahler (SK) geometry.  We now review the features of SK geometry essential for our argument; a more systematic description of SK geometry can be found in \cite{Freed:1997dp}, or in a more physical language in \cite{paper1, Argyres:2018zay}.

Dirac quantization implies that the electric and magnetic (EM) $\U(1)^r$ charges of states lie in a rank $2r$ lattice, $\bp\in\Z^{2r}$, equipped with an integral, skew-symmetric, and non-degenerate Dirac pairing, $\vev{\bp,\bq}\in\Z$.  The EM duality group, $\SpDrZ$, is the discrete group of basis changes of this lattice preserving the Dirac pairing.

An SK geometry is specified by its \emph{SK section}, $\bs(\bu)\in\C^{2r}$, which is locally a holomorphic vector field on the CB whose components are the $r$ special coordinates and the $r$ dual special coordinates on the CB.  More precisely, $\bs(\bu)$ takes its values in $V^*$, the linear dual of the complexification of the charge lattice, $V=\C\otimes\Z^{2r}\simeq\C^{2r}$, which inherits a symplectic structure from the Dirac pairing, as well as a linear action of the EM duality group.  
%We will call $V$ the charge space and $V^*$ the dual charge space.  
$\bs$ is a holomorphic section of an $\SpDrZ$ bundle: upon continuing it around a closed path $\g$ in the CB it may be multiplied by an element $M_\g$ of the EM duality group, called the EM duality monodromy around $\g$.  $\bs$ satisfies the SK integrability condition $0 = \vev{\del_j\bs, \del_k\bs}$ for all $j,k\in\{1,\ldots,r\}$ where $\del_j = \del/\del u^j$ are derivatives with respect to the CB complex coordinates.

The CB geometry is constructed from $\bs$.  The CB K\"ahler potential is $K = i \vev{\bs, \bar\bs}$, and hence the CB hermitean metric is $ds^2 = i \vev{\del_j\bs, \delb_\kb\bar\bs}\, du^j\, d\bar u^\kb$.  The central charge $Z_\bp$ in the EM charge sector $\bp$ of the IR effective theory at a vacuum $\bu$ is given by $Z_\bp(\bu)=\bp^T \bs(\bu)$.  The central charge computes the mass of BPS states, so $\bs$ has mass dimension $\D_\bs=1$, providing the basic normalization of CB scaling dimensions.

Finally, CBs of $\cN=2$ SCFTs have some special properties.  The dilatation and $\U(1)_R$ generators of the $\cN=2$ superconformal symmetry algebra act non-trivially on the CB since these symmetries are spontaneously broken by all but the unique conformal vacuum, $\cO\in$ CB.  ``CB operators" are the scalar chiral conformal primary operators, $\Phi_a$, of the SCFT which can get non-zero vevs on the CB.  These are the $\cE^r_{(0,0)}$ operators in the nomenclature of \cite{Dolan:2002zh}.  Because scaling dimensions and $\U(1)_R$ charges of CB operators are equal, the two symmetries together act by complex scalings
\begin{align}\label{cplxscal}
\Phi_a \mapsto \l^{\D_a} \Phi_a,
\qquad \l\in\tCs,
\end{align}
where $\D_a$ is the real positive scaling dimension of $\Phi_a$ and $\tCs$ is the infinite-sheeted Riemann surface of $y=\ln x$ covering the punctured complex plane, $x\in\C^*$.

\subsubsection{Physical metric singularities of CB geometries} 

CB geometries are SK geometries with singularities (non-analyticities).  If there is a BPS state in the spectrum of the theory with charge $\bp$ at a point in the CB where $Z_\bp(\bu)=0$, then there will be a massless charged state at that vacuum.  In that case the  description of the CB effective action in terms of free vector multiplets breaks down.  In particular, along the locus where $Z_\bp(\bu)=0$ 
%and if a BPS state with charge $\bp$ is in the spectrum, 
the CB metric may be non-analytic, and the SK structure (indeed, the complex manifold structure) of the CB may be ill-defined.  Denote by $\cV$ the set of all points with such singularities in the CB SK geometry.

We will review here the result of requiring physical consistency of the CB effective action in the vicinity of $\cV$ together with certain regularity assumptions, discussed in sections 2.2 and 4.2 of \cite{Argyres:2018zay} for rank 2 CBs.  In all cases the arguments given there either apply directly to the general-rank case or are easily generalized.  The main result is that with these assumptions the SK structure can be continued through $\cV$ in a precise sense and that EM duality monodromies of $\bs$ make sense for paths which intersect $\cV$.

We \emph{assume} that the underlying CB complex geometry is $\C^r$, which is equivalent to assuming that the reduced CB chiral ring is freely generated; see \cite{Argyres:2017tmj} for a critical discussion of this assumption.  This gives a preferred set of CB coordinates $\bu = (u^1, \ldots, u^r) \in \C^r$ where $u^k = \vev{\F_k}$ are the vevs of a generating set of CB operators of definite scaling dimensions, $\D_k$.  This basis diagonalizes the complex scaling action on the CB, 
\begin{align}\label{Csact}
\bu \mapsto  \l \circ \bu = 
\left( \l^{\D_1} u^1 , \ldots, \l^{\D_r} u^r \right) 
\qquad \l \in \til\C^*.
\end{align}
$\U(1)_R$ transformations correspond to this action with $\l= \exp(2\pi i\vf)$ a phase, and will play a central role in our discussion.  Unitarity of the SCFT implies $\D_k\ge1$.
 
The SK section $\bs$ does not diverge anywhere on the CB, for otherwise there would be a subsector of the theory which decoupled at all scales \cite{Argyres:2018zay}.  $\cV$ is closed in CB, since otherwise there is no consistent physical  interpretation of its IR effective action at the boundary points of $\cV$ (which are not in $\cV$).  The CB metric is complete and positive definite; in particular $\cV$ is at finite distance and distances along paths within $\cV$ are well-defined.  $\cV$ is a union of components each defined by the vanishing of a central charge $Z_\bp$ for some $\bp$.  $Z_\bp$ is a locally holomorphic function but is non-analytic at its zeros, where it has branch points.  

We further \emph{assume} that $\cV$ is a complex analytic set in the CB.  This assumption is made to avoid the possibility of having a closed $\cV$ which has accumulation points in its transverse complex plane or even a continuum of such points making it effectively have real codimension 1 in the CB.  It seems possible that this assumption actually follows from the local holomorphicity of $\bs$ together with the fact that there are only a countably infinite number of possible central charges (labeled by the charge lattice) whose vanishing can define $\cV$.

In any case, it then follows that $\cV$ is a complex codimension 1 variety in the CB, and a generic point of $\cV$ is a regular complex hypersurface in the CB.  From this and the metric regularity properties, it follows that $\bs$ has a well-defined limit as one approaches almost every point of $\cV$ from a transverse direction.  And though the transverse derivatives of $\bs$ generally diverge at $\cV$, the limit of the tangential derivatives of $\bs$ at $\cV$ are well defined and non-vanshing.

In \cite{Argyres:2018zay} more detailed restrictions on the behavior of $\bs$ were derived.  In particular the EM duality monodromy around a small loop linking a single component of $\cV$ defined by the vanishing of the central charge $Z_\bp = \bp^T\bs$ was shown to have a simple factorized form, from which it follows that its linking monodromy, $M_\cV$, is unipotent.  From this it follows that there exists a polynomial $\msP$ such that $f_\S := \bp^T \msP(M_\cV) \bs$ is a (single-valued) homorphic function vanishing along $\cV$.  This explicitly realizes $\cV$ as an analytic set in the CB in terms of the non-analytic SK section $\bs$ and its linking monodromy.  This provides the fundamental link between the topological and complex-analytic data from which the explicit CB geometry can be constructed by analytic continuation.  We will not use this here, except in passing in a decoupling argument in the next subsection, so will not elaborate on it further.

\subsubsection{Commensurateness of CB scaling dimensions}

Closedness of $\cV$ and complex scale invariance imply that the spectrum $\msD := \{\D_1,\ldots,\D_r\}$ of CB dimensions must be commensurate, i.e., $\D_j/\D_k\in\Q$ for all $j$, $k$.  This argument was also given in the rank 2 case in \cite{Argyres:2018zay}, but since it is qualitatively different in the higher rank case, we give the general argument in more detail.  We will argue this by showing that if they are not commensurate then in order for the locus of metric singularities to be closed in the CB, the CB geometry must factorize into the direct product of lower rank scale-invariant CB geometries.  Then if follows by induction that all CB dimensions are commensurate since it is known that they are all rational in the rank 1 case.

Partition the set of CB coordinates into subsets whose dimensions are commensurate: $\bu = \bu^{(1)} \amalg \cdots \amalg \bu^{(p)}$ with $\bu^{(a)} := \{ u_{(a)}^1, \ldots, u_{(a)}^{r_a}\}$ with $\sum_a r_a = r$.  The corresponding partition of their set of dimensions is $\msD = \msD_1 \amalg \cdots \amalg \msD_p$ where each subset is $\msD_a = \{ \D^{(a)}_1 , \ldots \D^{(a)}_{r_a}\}$ with $\D^{(a)}_j/\D^{(a)}_k\in\Q$ but $\D^{(a)}_j/\D^{(b)}_k\notin\Q$ for all $a\neq b$.

Consider a neighborhood of a regular point in the complex structure of the locus of metric singularities where it is locally cut out by a holomorphic function, $\cV := \{\bu | f(\bu)=0\}$, and consider a non-singular point $\bu_*\notin\cV$ in this neighborhood all of whose scaling coordinates are non-vanishing.

Complex scale invariance implies $f(\l\circ\bu_*) = \l^{\D_f} f(\bu_*)$.  The scaling dimension of $f$ is positive, $\D_f>0$, since it is holomorphic in the $\bu_*$ which have positive dimensions, and is not constant.  The derivative of the scaling relation for $f$ with respect to $\l$ evaluated at $\l=1$ implies $\del_\bl f=\D_f f$, where $\del_\bl$ is the derivative along the tangent vector $\bl$ to the $\l\circ\bu_*$ orbit at $\l=1$; in components $\ell^j := \D_j u_*^j$.

In addition to this consequence of scaling under excursions in a small connected neighborhood of $\l=1$, one can also consider large excursions in the phase of $\l$ so that $\l$ is again in an arbitrarily small neighborhood of 1 and $\l\circ\bu_* = \bu_* + \ve\bv$ with $\ve\in\C$ arbitrarily small.  Under such an excursion, if all the scaling dimensions are commensurate, then $\bv \propto\bl$.  But if $\msD$ has $p$ mutually incommensurate subsets, then the image of $\l\circ\bu_*$ for arbitrary phase of $\l$ is dense in a $p$-dimensional submanifold of the CB through $\bu_*$.  The tangent space to this submanifold is the subspace of the tangent space to the CB at $\bu_*$ spanned by $\bl^{(a)}$, $a=1,\ldots, p$.  Here $\bl^{(a)}$ has components only in the subspace of the tangent space spanned by derivatives with respect to the $\bu^{(a)}$ commensurate subset of the CB coordinates, and is proportional to $\bl$ in that subspace; in components $(\ell^{(a)})_j = \D^{(a)}_j u_{(a)*}^j$.  

The denseness of the image of $\l\circ\bu_*$ in this $p$-dimensional submanifold through $\bu_*$ implies that an $N\in\Z$ exists such that for $\l=e^{2\pi i N}(1 + \ve) \in\tCs$ we have $\l\circ\bu_* \approx \bu_* + (\ve /\d_a) \bl^{(a)}$ for any $a$ and $\ve$ to any desired accuracy.   Here $\d_a = \gcd\msD_a/\sum_b \gcd\msD_b\in (0,1)$ is different for each $a$ by the mutual incommensurateness of the $\msD_a$ sets.   Putting this in the complex scaling relation for $f$ and expanding to first order in $\ve$ then implies $\del_{\bl^{(a)}}f=\d_a\D_f f$ for each $a$.  This means that $f$ is separately homogeneous in each commensurate block with a different scaling weight.  This is only possible if $f = \prod_{a=1}^p f_a(\bu^{(a)})$ where each factor is a function of the scaling variables in only a single block.  (This follows, for instance, by taking derivatives of the block scaling relations to find $\del_{\bu^{(b)}}\del_{\bl^{(a)}} \ln f = 0$ for all $a\neq b$.)  Since this is true for generic $\bu$ and $\bu$ are globally defined coordinates on the CB, it follows that $f$ factorizes, and $\cV$ is the union of varieties which are each extended in all but one commensurate block $\bu^{(a)}$ of CB scaling coordinates.  

As mentioned at the end of the last subsection, the analytic form of $f$ near $\cV$ provide the boundary conditions for the analytic continuation of $\bs$ to the whole CB, thereby defining the CB geometry.  Since $f$ factorizes as above, it follows that so does $\bs$ and therefore also the CB geometry is a direct product of lower-rank scale invariant CB geometries.  (To reach this conclusion also requires an argument showing that the dual charge space $V^*$ in which $\bs$ takes its values also decomposes into a direct sum of symplectic subspaces corresponding to each commensurate coordinate block;  this argument is given in the rank-2 case in \cite{Argyres:2018zay} and generalizes immediately to the general rank case.)  Induction in the rank then implies that all CB scaling dimensions are commensurate, concluding the argument.

\subsection{U(1)$_{\bf R}$ monodromies}

$\U(1)_R$ monodromies \cite{Argyres:2018zay} are $\bs(\bu)$ monodromies associated to analytic continuation along $\U(1)_R$ orbits. 
%Following the notation introduced in \cite{Argyres:2018zay}, we will denote these monodromies with the fancy script $\msM$.
The $\U(1)_R$ orbit of a generic point $\bu_*$ with $r$ non-vanishing entries is the path obtained by specializing the $\tCs$ action \eqref{Csact} on $\bu_*$  to a pure phase, $\l= \exp(2\pi i\vf)$:
\begin{eqnarray}\label{U1Ror}
&\g_* :=\{\, u^j =  u^j_* e^{i\D_j\vf}, \ \vf\in[0,2\pi/S)\},\\\nonumber
&S \in \R^+,\quad \D_j = S \cdot \ell_j ,\quad \ell_j\in\N, 
\quad \gcd( \ell_1,\ldots, \ell_r)=1.
\end{eqnarray}
$S$ is well-defined and the path is closed because the $\D_j$ are all commensurate.  
Denote by $\msM_* \in \Sp(2r,\Z)$ the EM duality monodromy around $\g_*$.  Since the SK section has dimension 1, $\bs( \l\circ \bu) =  \l \, \bs(u)$, and the $\msM_*$ monodromy satisfies
\beq\label{unitnorm}
\msM_*\, \bs(\bu_*) = \exp(2\pi i/ S)\  \bs(\bu_*)
:=\m_*\, \bs(\bu_*).
\eeq
Therefore $\msM_* \in \SpDrZ$ has a unit-norm eigenvalue, $\m_*$.  But $\msM_*$ and $\m_*$ cannot vary as $\bu$ varies since $\SpDrZ$ is discrete.  Derivatives of \eqref{unitnorm} gives $\msM_*\,\del_j \s(\bu)  =  \m_*\ \del_j \s(\bu)$ which, by the integrability conditions, implies that the $\m_*$ eigenspace contains an $r$-complex dimensional lagrangian subspace of $\C^{2r}$.  A basic property of symplectic linear algebra is that $\Sp(2r)$ eigenvalues occur in reciprocal pairs whose joint generalized eigenspace is symplectic (see e.g., \cite{Freitas:2004}), therefore 
\begin{result} \label{alleigen}
\emph{All} eigenvalues of $\msM_*$ have unit norm for generic $\bu_*\in$ CB.
\end{result}
\noindent For a given rank $r$, there is a finite and computable set of allowed eigenvalues for such matrices.  An important property of this set, derived below, is that they are all of the form $\mu_*=\exp(i2\pi m/n)$ with $m,n\in\Z$.  Notice that determining possible unit-norm eigenvalues of $\msM_*$ does not directly determine the allowed $\D_j$, but only their greatest common divisor, $S$.  But this is enough to show that $S\in\Q$ implying that all the $\D_j$ are not just commensurate but rational.

Result \ref{alleigen} is true regardless of the structure of the singularities.  In fact as explained at the beginning of the previous section, the SK section can be continued to the singular subvariety where it has a well-defined value.  In particular, it has well-defined EM duality monodromies along paths which intersect the singularities. 

\subsection{Rank $r$ scaling dimensions}

The $\U(1)_R$ orbits $\g_*(\bu)$ do not vary continuously (are not homotopic) as $ \bu_*$ is varied through special values.  For example, consider the orbit $\g_1$ through $\bu^1_*$ with coordinates $(u_*^1, \ldots, u_*^r) = (1, 0, \ldots, 0)$ 
\begin{align}
\g_1 = \{ \bu \ |\ u^j = \d^j_1 e^{i \D_1 \vf}
\ \text{for}\ \vf\in[0,2\pi/\D_1)\ \}.
\end{align}
The subscript (superscript) in $\g_1$ ($\bu^1_*$) denotes the position of the only non-vanishing entry.  Call $\msM_1$ the monodromy around $\g_1$.  The analog of \eqref{unitnorm} for $\msM_1$ reads
\begin{align}\label{g1monod}
\msM_1\, \bs(\bu^1_*) = \exp(2\pi i/ \D_1)\  \bs(\bu^1_*).
\end{align}
Comparing this with $\g_*$ in \eqref{U1Ror}, we obtain $\g_* \sim (\g_1)^{\ell_1}$ (homotopic).  Using the fact that for (EM) monodromies $M_\g = M_{\g'}$ if $\g$ is homotopic to $\g'$, this implies
\beq
\msM_{*} = (\msM_1)^{\ell_1} .
\eeq
Because of Result \ref{alleigen} we conclude that $\msM_1$ also has all unit-norm eigenvalues.  The discussion above generalizes to all $\msM_j$, $j\in\{1,\ldots,r\}$, thus we can conclude that
\begin{result}
\emph{All} eigenvalues of $\msM_j$, $j=1,\ldots,r$, have unit norm. 
\end{result}
\noindent Determining the unit norm eigenvalues of the $\msM_j\in \Sp(2r,\Z)$ now directly allows one to compute the set of allowed $\D_j$.  

\subsubsection{Unit norm eigenvalues of unipotent $\SpDrZ$ elements}\label{Sp2r}

We find the allowed eigenvalues of $\SpDrZ$ matrices with all unit-norm eigenvalues by classifying their possible characteristic polynomials, $\msP_{2r}(\m)$.  $\msP_{2r}$ has degree $2r$, integer coefficients, and all its roots are roots of unity.  The irreducible integer-coefficient polynomials whose roots are roots of unity are the cyclotomic polynomials, $\msC_n$:
\begin{align}\label{cyclopoly}
\msC_n(\m) &= \!\!\! \prod_{\substack{0< m \le n\\\gcd(m,n)=1}} 
\!\! \left( \m-e^{i2\pi m/n}\right)\, ,
\\
\text{degree}(\msC_n) &= \ \vf(n)\  
= \ n \textstyle\prod_{p|n} (1-p^{-1})
\nonumber
\end{align}
where $\vf(n)$ is the Euler totient function which counts the primitive $n$th roots of unity.  By the definition of the $\msC_n$, the characteristic polynomial can be written as
\begin{align}\label{chapoly}
\msP_{2r} = \textstyle\prod_{j=1}^N  (\msC_{n_j})^{a_j}
\end{align}
such that $\sum_{j=1}^N a_j\,\varphi(n_j)=2r$.  
%The $a_j$ labels the multiplicity of each irreducible factor.  
A matrix whose characteristic polynomial is given by \eqref{chapoly} will be indicated as $[n_1^{a_1}\ldots n_N^{a_N}]$.  We will also use the notation $[\ldots]^\ell$ to mean the characteristic polynomial of the $\ell$th power of the matrix with characteristic polynomial [$\ldots$].   
%It is important to stress that while the set of eigenvalues is an invariant of $\SpDrZ$ conjugacy classes, it is a coarser characterization: there could be many different distinct conjugacy classes sharing the same set of eigenvalues. 

It follows directly from \eqref{cyclopoly} and \eqref{chapoly} that eigenvalues all have the form $\exp(i2\pi m/n)$ with $\vf(n)\le2r$, and comparing to \eqref{g1monod} and imposing the unitarity condition $\D_j\ge1$, we obtain our {\bf main result:}
\begin{align}\label{mainres}
\D_j \in \biggl\{\, \frac{n}{m} 
\ \bigg\vert 
\ \vf(n){\le}2r, 
\ 0{<}m{\le}n, 
\ \gcd(m,n){=}1
\, \biggr\}.
%\nonumber
\end{align}
This agrees with the result of \cite{Caorsi:2018zsq} who explicitly list the allowed values for low ranks in their table 8.
%Computing the allowed values for the lowest ranks, our results perfectly match what reported in table 8 of \cite{Caorsi:2018zsq}. 
Notice that if $\D_i=n_i/m_i$ then the characteristic polynomial of $\msM_i$ has to have at least one $\msC_{n_i}$ factor, i.e. be of the form $[\ldots n_i^{a_i}\ldots]$.

\subsection{Constraints on $r$-tuples}

Not all $r$-tuples of dimensions with entries in \eqref{mainres} are allowed.  A similar phenomenon was already pointed out in the rank-2 case \cite{Argyres:2018zay}.  Here we outline a set of relatively strong constraints on the possible $r$-tuples.

Consider $\{\D_1,\ldots,\D_r\}$ and decompose them as in \eqref{U1Ror} in terms of mutually prime $\ell_j$'s and a common rational factor $S$.  A first strong constraint comes from the fact that $\msM_*$, the monodromy which determines $S$, has an $r$-dimensional lagrangian subspace and its characteristic polynomial can be one of only 5 possibilities: [$1^{2r}$], [$2^{2r}$], [$3^r$], [$4^r$], or [$6^r$].   This follows from basic properties of symplectic matrices: if a unit-norm eigenvalue $\m$ has a (generalized) eigenspace of dimension $d$ which is isotropic (i.e., the symplectic product of all its eigenvectors vanishes), then the conjugate eigenvalue $\bar\m$ also has a $d$-dimensional isotropic eigenspace.  Thus a necessary condition for the eigenspace of $\m$ to contain an $d$-dimensional isotropic space is that $\m$ has at least multiplicity $d$ if $\m\neq\bar\m$ and multiplicity $2d$ if $\m=\pm1$.  As $d$ increases this becomes a progressively more constraining condition.  For $d=r$, and $\m\neq\pm1$, $\m$ and $\bar\m$ are the only eigenvalues and thus the characteristic polynomial has the form $[n^r]$.  But given that the order of $\msP_{2r}$ is $2r$ it follows that $\vf(n)=2$, which readily implies $n=3,4,6$. Similarly, if $\m=\pm1$ then the characteristic polynomial is just $[1^{2r}]$ or $[2^{2r}]$.

Thus the mutually prime $\ell_j$'s have to be such that $[\ldots n_k^{a_k}\ldots]^{\ell_j}=[\#^{r\ \text{or}\ 2r}]$ for all $j=1,\ldots,r$.  This is remarkably constraining.  For instance consider an $r$-tuple with all entries equal to a common scaling dimension $\bar\D$.  This implies that all the $\ell_j=1$ and in turn all the $\msM_j$'s must themselves have an eigenspace containing an $r$-dimensional lagrangian subspace.  From this it follows that $\bar\D$ can only take the 8 values allowed for the rank-1 case.

Consider the $\U(1)_R$ orbit of a base point with one of its components vanishing, say $u^i_*=0$.  Denote the vector of complex coordinates of the base point as $\bu_*^{\hat i}$, the associated $\U(1)_R$ orbit as $\g_{\, \hat i}$, and the monodromy around it as $\msM_{\hat i}$.   Since any vanishing entry of the base point $\bu_*$ is fixed by the $\U(1)_R$ action,  $\msM_{\hat i}\, \bs(\bu^{\hat i}_*) = \exp(2\pi i/ S_{\, \hat i}) \bs(\bu^{\hat i}_*) := \mu_{\, \hat i}\, \bs(\bu^{\hat i}_*)$, where $S_{\, \hat i}=S\cdot{\rm gcd}(\ell_1,\ldots,\ell_{i-1},\ell_{i+1},\ldots,\ell_r):=S\cdot{\rm gcd}_{\, \hat i}(\ell_1,\ldots\ell_r)$ is the common factor of all the $\D$'s omitting $\D_i$.  This immediately implies a set of homotopy equivalences among the $\U(1)_R$ orbits, and so the relates their monodromies:  $\msM_{\hat i}=\msM_j^{\ell_j/{\rm gcd}_{\hat i}(\ell_1,\ldots,\ell_r)}$.  Furthermore, analogously to the argument giving result \ref{alleigen} for $\msM_*$, since we can continuously deform $\g_{\, \hat i}$ in the $u^i=0$ hyperplane, the eigenspace of $\m_{\, \hat i}$ must contain an $(r-1)$-dimensional isotropic subspace.

Now consider $\U(1)_R$ orbits of base points with an increasing number of vanishing components.  The argument above generalizes straightforwardly.  In this way we derive a nested set of constraints which restricts the set of allowed $r$-tuples.  We have not attempted to compute the resulting set of allowed $r$-tuples for general rank.  It would be interesting to compare the result obtained imposing the conditions described here with tables 9 and 10 of \cite{Caorsi:2018zsq}.

\begin{figure}
\includegraphics{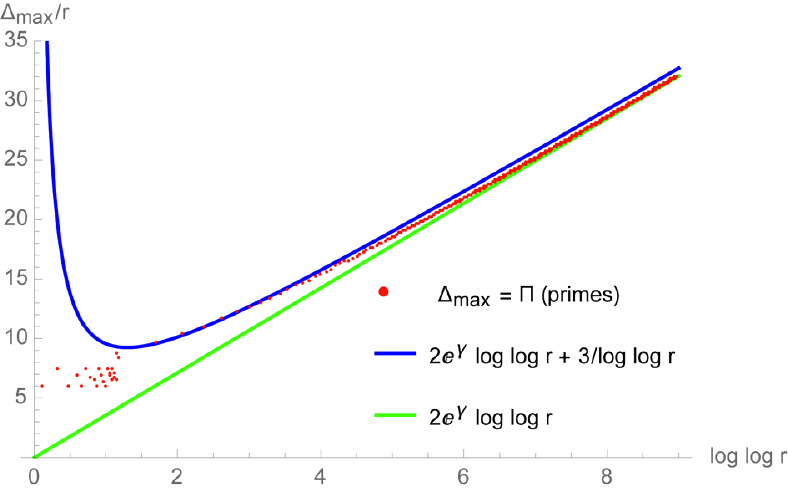}
\caption{\label{fig1} The ratio $\D_\text{max}/r$ is plotted in red against $\ln\ln r$.  The solid curves are two asymptotic limits bracketing $\D_\text{max}/r$ for comparison.}
\end{figure}

\subsection{Large rank limit}

One notable property of our main result is that the maximum CB dimension grows superlinearly with the rank:
\begin{align}
\text{max}( \D_j) &= \text{max}\{n \, | \, \vf(n){\le}2r \} \ \sim_{r\to\infty}\  2 r \, e^\g \ln\ln r,
\nonumber
\end{align}
where $\g\approx 0.577$ is the Euler-Mascheroni constant.  This asymptotic growth follows from inverting the lower bound $n/\vf(n) < e^\g \ln\ln n + 2.51 /\ln\ln n$ for $n>2$ and the fact that $e^\g\ln\ln n < n/\vf(n)$ for infinitely many $n$ \cite{RS62}.  Indeed, it is easy to see from the expression for the totient function, $\vf(n) = n \prod_{p|n}(1-p^{-1})$, where the product is over all primes dividing $n$, that locally maximal values of $n/\vf(n)$ occur for $n$ a product of the first $N$ primes.  These correspond to maximal dimensions $\D_\text{max} = \prod_{i=1}^N p_i$ at rank $r=\frac12 \prod_{i=1}^N (p_i-1)$.  Figure \ref{fig1} plots $\D_\text{max}/r$ versus $\ln\ln r$ for these values in red as well as for $3\le r \le 30$.

The superlinear growth of $\D_\text{max}$ with rank raises the possibility of series of $\cN=2$ SCFTs with well-defined large-$r$ limits which include such $\D_\text{max}(r)$ in their spectra of CB dimensions.  A well-defined large-$r$ limit presumably means a flavor symmetry independent of $r$ as $r\to\infty$, and leads to the expectation that $a - c \sim c \,\cO(1/r)$ in that limit, where $a$ and $c$ are the 4d conformal central charges.  The central charge sum rule \cite{st08} relating them to the spectrum of CB dimensions via $2a-c = \frac14 \sum_{j=1}^r (2\D_j-1)$ then leads to the expectation that $a \sim c \sim r^2 \ln\ln r$ for an ``evenly spaced" spectrum of CB dimensions.

The holographic dual of such a (hypothetical) series would presumably be a warped product of AdS$_5$ with a compact 5-manifold supporting $\cO(r)$ units of 5-form flux.  It is an interesting question whether such a geometry exists with $c \sim r^2 \ln\ln r$.  The 4d holographic duals of series of $\cN=2$ SCFTs that the authors are aware of all have $c \lesssim r^2$. Note that warped AdS duals of series of theories in various dimensions have been constructed in which there are logarithmic corrections to power-law large-$r$ scaling of central charges or sphere partition functions, but these theories do not have stable flavor groups at large $r$ --- e.g., \cite{Assel:2011xz, Assel:2012cp} --- or are not SCFTs --- e.g., \cite{Aharony:2005zr}.

\acknowledgments

It is a pleasure to thank A. Buchel, J. Distler, C. Long, M. Lotito, E. Perlmutter and C. Uhlemann for useful discussions.  PCA was supported in part by DOE grant DE-SC0011784 and by Simons Foundation Fellowship 506770.  MM was supported in part by NSF grant PHY-1151392 and in part by NSF grant PHY-1620610.

\end{document}